\documentstyle[12pt]{article}
\begin{document}
\title{Exact Ground State Energy of the Strong-Coupling Polaron}

\author{Elliott~H.~Lieb$^1$ and Lawrence~E.~Thomas$^2$\\
\footnotesize \it $^1$Departments of Physics and Mathematics, 
Jadwin Hall, Princeton University,\\
\footnotesize \it P.~O.~Box 708, Princeton, New Jersey 08544\\ 
\footnotesize \it
$^2$Department of Mathematics, University of Virginia,
Charlottesville, Virginia 22903}
\date{December 13, 1995, revised May 4, 1996, erratum inserted April
  28, 1997}
\maketitle

Dedicated to the memory of R.L.~Dobrushin and P.W.~Kasteleyn, two
founders of modern statistical mechanics.

\begin{abstract}

  The polaron has been of interest in condensed matter theory and
  field theory for about half a century, especially the limit of large
  coupling constant, $\alpha$.  It was not until 1983, however, that a
  proof of the asymptotic formula for the ground state energy was
  finally given by using difficult arguments involving the large
  deviation theory of path integrals.  Here we derive the same
  asymptotic result, $E_0\sim -C\alpha^2$, and with explicit error
  bounds, by simple, rigorous methods applied directly to the
  Hamiltonian. Our method is easily generalizable to other settings,
  e.g., the excitonic and magnetic polarons.

\end{abstract}
\renewcommand{\baselinestretch}{1.5}


The polaron Hamiltonian of Fr\"ohlich \cite{HF} is a model for the
Coulomb interaction of one or more electrons with the quantized
phonons of an ionic crystal.  In the course of time it was also seen
to be an interesting field theory model of non-relativistic particles
interacting with a scalar boson field, and it was widely studied
\cite{P,Bog,Tyab,EG,JF,HS,LLP,Schultz} in both contexts.
\vfill
\footnoterule{\footnotesize \copyright 1996 by the authors.
  Reproduction of this article, by any means, is permitted for
  non-commercial purposes.}
\newpage 
The model has one dimensionless coupling constant, $\alpha$, and it
was noticed very early that there seems to be a qualitative difference
in the ground state between the weak coupling
regime, well described by perturbation
theory \cite{LLP}, in which the electron is spread out, andthe strong
coupling regime in which the electron appears to be trapped in a
phonon hole of its own making. (This question of trapping seems to
have first been considered by Landau \cite{Landau} in the context of a
classical phonon field.)

By now it seems doubtful that such a trapping actually occurs
\cite{HS,GL,PD}, but it remains true that the calculation of the
ground state energy, $E_0(\alpha)$, is very different in the two
regimes. The strong coupling theory was studied by Pekar \cite{P} (see
also \cite{Bog,Tyab,EG,Schultz}) who hypothesized that in this limit
the total ground state wave function $\Psi$ could be taken to be a
product of an electronic function $\phi(x)$ and a phonon function
$\vert {\bf \xi} \rangle$.  It is a fact that this ansatz, called
the adiabatic approximation, yields exactly the same result for the
ground state energy as the corresponding model with classical phonons.
This stems from the fact that the phonon operators enter only linearly and
quadratically in the Hamiltonian. 

Fr\"ohlich's Hamiltonian in appropriate units is
\begin{equation}\label{eq:1}
H= {\bf p}^2 + \sum_{\bf k}{a}_{\bf k}^*{a}_{\bf k}^{\phantom*} +
                (\frac{4\pi\alpha}{V})^{1/2}\sum_{\bf k}
\left[\frac{a_{\bf k}}{|{\bf k}|}
      e^{i{\bf k\cdot\bf x}} + h.c.\right]
\end{equation}
where ${\bf p}=-i{\bf \nabla}$ is the electron momentum operator,
${\bf x}$ is its coordinate, and $V$ is the volume of the crystal,
which tends to infinity.  The ${\bf k}$'s are the usual normal modes
( e.g., ${\bf k} = 2\pi V^{-1/3} (n_1, n_2,n_3)$ for a cubic box )
and, as usual, $V^{-1}\sum_{\bf k} \rightarrow (2\pi)^{-3}\int d^3k$.
Also, as usual, the ${\bf k} =0$ mode is omitted. 
When $\langle \Psi \vert H \vert \Psi \rangle$ is computed with
Pekar's ansatz one easily finds that the phonon part $\vert \xi
\rangle$ can be easily evaluated by \lq completing the square\rq\ and
the resulting $\phi $ minimizes the energy
\begin{equation}
{\cal E}(\phi) \equiv \int
\vert {\bf \nabla} \phi \vert^2 d^3x -\alpha \int \int \frac {\phi({\bf
x})^2\phi({\bf y})^2} {\vert {\bf x-y}\vert}d^3xd^3y\label{Pekar} 
\end{equation}
subject to $\int \phi({\bf x})^2 d^3x=1$.
This, as we mentioned above, is what one would get if the 
${a}_{\bf k}$'s are replaced by c-numbers $\xi_{\bf k}$.  

It has been proved \cite{EL} that there is exactly one $\phi$, up to
translations, that minimizes ${\cal E(\phi)}$. By scaling, the minimum
energy (call it $E_P(\alpha)$) is proportional to $-\alpha^2$. The
question we address is this: {\it Is this energy, $E_P(\alpha)$,
  asymptotically exact as $\alpha$ tends to infinity? } Since it is
given by a variational calculation, $E_P(\alpha)$ is an upper bound to
$E_0(\alpha)$; the problem is to find a lower bound that agrees with
$E_P(\alpha)$ to leading order. {\it We provide such a lower bound in
(\ref{E2}) below.}

For a long time more or less the only thing that could be done to
validate $E_P(\alpha)$ rigorously was a lower bound \cite{LY} of the right order,
$-\alpha^2$, but which was about a factor of 3 too large.  It was not
until 1983, with a precursor in \cite{AGL}, that a complete proof of
the asymptotic correctness of $E_P(\alpha)$ was
given by Donsker and Varadhan \cite{DV}. The proof starts with a
familiar expression for the ground state energy,
\begin{equation}\label{eq:f1}
  E_0(\alpha)=
  \lim_{\beta\rightarrow\infty}\frac{1}{\beta}\ln\langle\Psi|e^{-\beta
    H}|\Psi\rangle,
\end{equation}
with $\Psi$ such that its spectral resolution contains the ground
state energy or low energy spectrum of $H$ but is otherwise arbitrary.
Now in fact there is a relatively simple Feynman-Kac formula
for the semigroup $e^{-\beta H}$ first noted by Feynman in \cite{F};
by integrating out the phonon coordinates in this formula, one finds
that the right side of (\ref{eq:f1}) can be replaced by the expression
\begin{equation}\label{eq:f2}
  \lim_{\beta\rightarrow\infty}\frac{1}{\beta}\ln
  E\left\{\exp(\alpha\int_0^{\beta}\int_0^t \frac{e^{-(t-s)}}{|{\bf
      x}(t)-{\bf x}(s)|}ds
  dt)\right\}.
\end{equation}
Here, $E$ represents path space expectation for the Brownian motion
${\bf x}(t)$. The Feynman-Kac formula and functional integral are discussed
in some detail by Ginibre \cite{Ginibre} and Roepstorff \cite{Roep}. 

It is this limit (\ref{eq:f2}) that Donsker and Varadhan analyze, in
effect showing that the Pekar functional (\ref{Pekar}) provides a kind
of large deviation rate function for the functional integral in
(\ref{eq:f2}). Their proof is very far from simple and requires an
enormous knowledge of large deviation theory of Brownian motion, a
subject unknown to most physicists, and it does not generalize easily
to other models.  Owing to the double integral in (\ref{eq:f2}) it is
necessary to construct a large deviation theory for ``infinitely many
variables''. Moreover, no error bounds were provided \cite{V}.  We note that
previously Feynman had used this formula to estimate the ground state
energy from above, but his estimate is not sharp even for $\alpha\rightarrow\infty$.  See also \cite{Sa} where this functional
integral is again employed to obtain a lower bound on the ground state
energy which, however, is not sharp.

Regarding the matter of error bounds, we note that Gross \cite{EG} takes
advantage of momentum conservation to eliminate three phonon degrees of
freedom and transforms the Hamiltonian $H$ to a new form involving a
sum of $\alpha^2$-, $\alpha^0$- and $\alpha^{-2}$-terms. (He continues
to employ the Pekar ansatz.)  This form of the Hamiltonian perhaps
suggests that the ground state energy itself has an asymptotic
expansion in inverse powers of $\alpha^2$. But the terms in
this transformed Hamiltonian are unbounded operators, and so the order
of their contributions to the energy is by no means obvious. The
bound we obtain on the error is actually $O(\alpha^{9/5})$,
which is to be compared with the main term $O(\alpha^2)$. 

In any event, there is the following problem: The Pekar ansatz is
based on the physically appealing notion that at
large coupling the phonons cannot follow the rapidly moving electron
(as they do at weak coupling) and so resign themselves to interacting
only with the \lq mean\rq\ electronic density, $\phi({\bf x})^2$.
What, exactly, do these somewhat anthropomorphic words mean?  If they
are so very physical, they should be quantifiable and it should not
take four decades to build a proof of their correctness.  Moreover,
one should be able to find a proof that is relatively simple (since
the idea is a simple one), and one  that also yields
some kind of quantitative error estimate. It should also be robust
enough to allow an easy extension to some variations of $H$, such as
having several electrons instead of one (the polaronic exciton),
inclusion of magnetic fields, an electron kinetic energy, $T({\bf
  p})$,  other than ${\bf p}^2$  (for which the semigroup
$\exp[-T({\bf p})]$ might not even have a positive kernel), 
accommodation of ${\bf k}$-dependence and nonlinearity in the phonon
self energies, electron-phonon interaction energies other than
$1/\vert {\bf k} \vert$, etc.

The method presented here satisfies, we believe, the criteria of
simplicity, robustness and, of course, rigor. For this 
reason it might be of general interest
in condensed matter theory or field theory. Despite its generality we
shall, for clarity, restrict our discussion to the original
Hamiltonian, $H$. 

Before going into the details, it is helpful 
to give an
overview of our method. The first thing to notice is that Pekar's
ansatz amounts, mathematically, to saying that we can replace the
operators $a_{\bf k}$ by c-numbers $\xi_{\bf k}$. One might think, 
at first, that
such a replacement always leads to a lower bound for the energy, but
this is false. (If it were true, the small $\alpha$ energy would have to be
proportional to $-\alpha^2$ instead of the  much more negative, 
correct value $-\alpha$.) 
To pursue this idea, nevertheless,  the
natural tool to think of is coherent states for each phonon mode ${\bf k}$.
As is well known, these states are indexed by complex numbers $\xi =p+iq$ and, 
if $ \vert \xi \rangle \langle \xi \vert$ is
the projector onto the coherent state $\vert \xi \rangle$, 
we have that 
\begin{equation}
\int \vert \xi \rangle \langle \xi \vert \ d\xi d\xi^* =I,
\end{equation}
\begin{equation}
\int \xi \ \vert \xi \rangle \langle \xi \vert \ d\xi d\xi^* = a
\end{equation}
and 
\begin{equation}
\int (\vert \xi\vert^2 -1)\vert \xi \rangle \langle \xi \vert \ d\xi d\xi^*
= a^*a.
\end{equation}
{\it It is the extra term $-1$ in the last integral that kills the
  lower bound and gives an unwanted energy $-1$ for each phonon mode}
(and hence a negative contribution to the energy that is infinite both
because there is no infrared and no ultraviolet cutoff). In other
words, coherent states would give us what we want (effectively
replacing the operator $a$ by the number $\xi$), were it not for the
unfortunate fact that the positive {\it operator} $a^*a$ is
represented by the nonpositive {\it symbol}, $\vert \xi\vert^2 -1$.

Our remedy will be to reduce the effective number of phonon modes to a
{\it finite} number of $O(\alpha^{9/5})$, independent of $V$.
This 
will not sacrifice rigor because we shall prove that the reduction 
affects 
the energy only to $O(\alpha^{9/5})$ at most. These
modes will be quite different from the original $a_{\bf k}$ modes;
indeed they will be the $a_{\bf k}$ modes summed over boxes in ${\bf
k}$-space of size $\alpha^{3/5}$.  Thus, our physical description of
strong coupling will be a little different from the conventional one.
{\it Instead of saying that the phonons cannot follow the electron, our
point of view will be that the electron significantly excites only
finitely many field modes.}

This mode reduction is accomplished in several steps which can be
outlined as follows.

$\bullet$ {\bf I.} Using a simple commutator estimate (as in 
Schr\"{o}dinger's elementary
proof of Heisenberg's uncertainty principle \cite{Schr}) we can show that ${\bf \vert
k\vert}$ values larger than $K \approx \alpha^{6/5}$ can be ignored
with an energy cost of only $\alpha^{9/5}$.  This eliminates the
ultraviolet problem, i.e., the fact that  ${\bf |k|}^{-2}$ is not summable.

$\bullet$ {\bf II.} With the same energy error we can localize the electron
to a cube of side length $\alpha^{-9/10}$.

$\bullet$ {\bf III.}  We decompose the ball $\vert {\bf k}\vert <K$
into blocks of size $\alpha^{3/5}$, each containing $n$ ${\bf
k}$-values (with $n\sim V\alpha^{9/5}$). There are about
$\alpha^{9/5}$ blocks.  Because the electron coordinate ${\bf x}$ has
been localized, the function $\exp(i{\bf k\cdot x})$ in $H$ can be
replaced, in a block $B$, by $\exp(i{\bf k}_B\cdot {\bf x})$, where
${\bf k}_B$ is any conveniently chosen point in $B$
; The energy penalty is again $\alpha^{9/5}$. 
The effect of
this decoupling is that the electron interacts only with the one mode
$A_B \approx n^{-1/2}\sum_{{\bf k}\in B} a_{\bf k}$, from block $B$.
Thus, the electron ends up interacting with only $\alpha^{9/5}$ modes
, which is a finite number!  

$\bullet$ {\bf IV.} These active modes can now be represented by
coherent state integrals, as above. The unwanted $-1$ term now
contributes only $\alpha^{9/5}$ to $E_0$, that being the number of
modes.

The details will now be given.\\ 

{\bf I.} We first establish a commutator inequality which will show
 that the large ${\bf k}$ modes of the Hamiltonian may be discarded at
 the price of only a small decrease in the coefficient of the ${\bf
 p}^2$ term of the Hamiltonian.  {\it The argument given here is
 incorrect.  Please see the erratum at the end of the article.} For any normalized state with
 expectation $\langle\cdot\rangle$, we have
\begin{eqnarray}
        |k_j|\,|\langle{a}_{\bf k}e^{i{\bf k}\cdot{\bf x}}\rangle|=
  |\langle[p_j,{a}_{\bf k}e^{i{\bf k}\cdot{\bf x}}]\rangle|\nonumber\\
\leq 2\langle p_j^2\rangle^{1/2}\langle{a}_{\bf k}^*{a}_{\bf k}\rangle^{1/2}
\end{eqnarray}
for each phonon  momentum coordinate $k_j$, $j= 1,2,3$, just by the
Schwarz inequality.  Squaring this inequality, summing over $j$,  and
then taking the square root, we obtain \begin{equation}
        |\langle {a}_{\bf k}e^{i{\bf k}\cdot{\bf x}}\rangle|\leq
         \frac{2}{|{\bf k}|}\langle{\bf
         p}^{2}\rangle^{1/2}\langle{a}_{\bf k}^*{a}_{\bf
         k}\rangle^{1/2}.\label{modebound}
\end{equation}
A similar inequality holds for ${a}_{\bf k}$  replaced by ${a}_{\bf
k}^*$.  It follows that for any $\varepsilon>0$, and with $K\equiv
8\alpha/\pi \varepsilon$,
\begin{eqnarray}
\lefteqn{ -(\frac{4\pi\alpha}{V})^{1/2}\sum_{|{\bf k}|\geq
K}\left[\langle\frac{{a}_{\bf k}}{|{\bf k}|}e^{i{\bf k}\cdot{\bf
x}}\rangle+c.c.\right]}\nonumber\\ &\leq&
4(\frac{4\pi\alpha}{V})^{1/2}\langle{\bf p}^{2}\rangle^{1/2}
\sum_{|{\bf k}|\geq K}|{\bf k}|^{-2}\langle{a}_{\bf k}^*{a}_{\bf
k}\rangle^{1/2}\nonumber\\ &\leq&
4(\frac{4\pi\alpha}{V})^{1/2}\langle{\bf p}^2\rangle^{1/2}(\sum_{|{\bf
k}|\geq K} |{\bf k}|^{-4})^{1/2}(\sum_{|{\bf k}|\geq K}\!\!\!  \langle
\ {a}_{\bf k}^*{a}_{\bf k}^{\phantom *}\rangle)^{1/2}\nonumber\\  &=&
2\varepsilon^{1/2}\langle{\bf p}^2\rangle^{1/2}(\sum_{|{\bf k}|\geq K}
\langle \
{a}_{\bf k}^*{a}_{\bf k}{\phantom *}\rangle)^{1/2}\nonumber\\
&\leq& \varepsilon\langle{\bf p}^2\rangle +\sum_{|{\bf k}|\geq K} \ 
\langle \ {a}_{\bf k}^*{a}_{\bf k}{\phantom *}\rangle,\label{uvbound}
\end{eqnarray}
again by the Schwarz inequality. (We have taken the limit
$V\rightarrow \infty$ in the sum.)

The above inequality (\ref{uvbound}) is an ultraviolet bound.  It shows
that the Hamiltonian $H$ is bounded below by a new one $H_{K}$, i.e.,
$H\geq H_{K}$, with 
\begin{equation}
H_{K}\equiv(1-\frac{8\alpha}{\pi K}){\bf p}^2
+\sum_{|{\bf k}|<K}{a}_{\bf k}^*{a}_{\bf k}^{\phantom *} +
(\frac{4\pi\alpha}{V})^{1/2}
\sum_{|{\bf k}|<K}
\left[ \frac{{a}
 _{\bf k}}{|{\bf k}|}
      e^{i{\bf k}\cdot{\bf x}} + h.c. \right].
\end{equation}
By completing the square, we have that
\begin{equation} 
{a}_{\bf k}^*{a}_{\bf k}^{\phantom *}
+(\frac{4\pi\alpha}{V})^{1/2}\left[ \frac{{a}_{\bf k}}{|{\bf k}|}
e^{i{\bf k}\cdot{\bf x}} + h.c. \right] \geq -\frac{4\pi\alpha}{|{\bf
k}|^2V}
\label{modelowerbound} 
\end{equation} 
and if $K$ is chosen so that
the coefficient of ${\bf p}^2$ vanishes in $H_{K}$, i.e., $K=
8\alpha /\pi$, then
\begin{equation}
 H\geq H_{K}\geq -\frac{4\pi\alpha}{ V} \sum_{|{\bf k}|<K} |{\bf
 k}|^{-2}= -\frac{16}{\pi^2}\alpha^2,
\end{equation}
showing that $H$ is indeed  bounded below by $O(\alpha^2)$.  But to
obtain the sharp lower bound, we will take a  much  
larger $K$, namely $K= (8/\pi
c_1)\alpha^{6/5}$.  Here and below,  the $c_i$'s are constants independent
of $\alpha$.\\

{\bf II.} The next step is to localize the electron.  Although a-priori
the electron is in the box of volume V, it actually can be confined to a
box of much smaller size, with only a slight relative increase in its
energy.  More precisely, we have the following: Let $\Psi$ be any
normalized state of the electron and phonons and define
$E\equiv \langle\Psi|H_{K}|\Psi\rangle$.  Then, given $\Delta E>0$,
there exists a function $\phi$  of the electron coordinate ${\bf
x}$ alone (but depending on $\Psi$), with support in some 
cube of side length
 $L= \pi(3/\Delta E)^{1/2}$
such that
\begin{equation}
\langle\phi\Psi|H_{K}|\phi\Psi\rangle \!
\Bigm/  \! \langle\phi\Psi|\phi\Psi\rangle\leq E+\Delta E.
\end{equation}

To see this, let $\phi({\bf x})=\prod_{j=1}^3\cos((\Delta
E/3)^{1/2}x_j)$ inside the cube of side length $\pi (3/\Delta E)^{1/2}$
centered at the origin and $\phi= 0$ outside the cube.  Let  $\phi_{\bf
y}({\bf x})= \phi({\bf x}-{\bf y})$ and consider the integral
\begin{eqnarray}
        \int\!\!\!&\Big(&\!\!\langle\phi_{\bf y}\Psi|H_{K}|\phi_{\bf
        y}\Psi\rangle 
-(E+
\Delta E)\langle\phi_{\bf y}\Psi|\phi_{\bf y}\Psi\rangle\Big)d^3
y\nonumber\\
 & =&\int(|\nabla\phi|^2-\Delta E|\phi|^2) \ d^3 y = 0.\label{local}
\end{eqnarray}
(Note here that the cross terms $\nabla\phi_{\bf y} \nabla\Psi$
vanish after the $y$-integration.)   
Evidently, there must be a point
${\bf y}$ such that at this point, the integrand on the left side of
(\ref{local}) is non-positive (and where $\langle\phi_{\bf
y}\Psi|\phi_{\bf y}\Psi\rangle$ is non-zero).   This $\phi_{\bf
y}$ is the  $\phi$ that we need. The electron localization is now complete.

For our purposes, we take $\Delta E= c_2\alpha^{9/5}$, and the
assertion above together with the ultraviolet bound then implies that the
ground state energy $E_{0}$ of $H$ satisfies 
\begin{equation}
        E_{0}\geq {\inf_{\Psi}}\,
        ^\prime\langle\Psi|H_{K}|\Psi\rangle-c_2\alpha^{9/5}\label{Eo}
\end{equation}
where the infimum is taken over all normalized $\Psi$'s but having
their ${\bf x}$-support in a cube  of side length no larger than $L=
(3/c_2)^{1/2}\pi\alpha^{-9/10}$ somewhere in the large volume $V$.\\

{\bf III.} The next step is to group the phonon modes together into blocks,
which we take to be cubes with sides of length $P= c_3\alpha^{3/5}$ or, 
more precisely, the portion of those cubes lying in the
big ball $\{{\bf k}:|{\bf k}|<K\}$.   Let  $B$ be a block (cube)
of momenta with sides of length $P= c_3\alpha^{3/5}$ and let ${\bf
k}_B$ be {\it any  fixed point} within this block 
whose precise position 
will be determined later.  It is clear that the set of momenta 
$\{{\bf k}: |{\bf k}|<K\}$ can be covered by 
$N\equiv\frac{4}{3}\pi
K^3/P^3+lower\,\, order=(2^{11}/(3\pi^2
c_1^3c_3^3))\alpha^{9/5}+lower\,\, order$ such blocks. 

Assume that ${\bf x}$ itself varies over a cube of side length no
bigger than $L$ with center at a point ${\bf x}_0$ which we take to be
the origin ${\bf 0}$.  (If ${\bf x}_0$ is not the origin, phase
factors $\exp{(\pm i({\bf k}-{\bf k}_B)\cdot{\bf x}_0)}$ standing
before $a_{\bf k}$ and $a_{\bf k}^*$ in the Hamiltonian
$H_{K}$ are readily eliminated by a unitary transformation involving
the phonon variables only.) We have that for ${\bf k}\in B$,
\begin{equation}
|e^{i{\bf k}\cdot{\bf x}}-e^{i{\bf k}_B\cdot{\bf x}}|
\leq|({\bf k}-{\bf k}_B)
\cdot{\bf x}|
\leq \frac{3\pi c_3}{2}(\frac{3}{c_2})^{1/2}\alpha^{-3/10}.
\label{exponentialdiff}
\end{equation}
Then, for any $\delta>0$, we have the 
following inequality, which is obtained 
by completing the square, as in (\ref{modelowerbound}). 
\begin{eqnarray}
\lefteqn{\!\!\!\!\!\!\!\!\!\!\!\!\!\!\!
\sum_B\sum_{{\bf k}\in B}\left[\delta{a}_{\bf k}^*{a}_{\bf k}^{\phantom
*} +(\frac{4\pi\alpha}{V})^{1/2}\big[\frac{{a}_{\bf k}}{|{\bf k}|}(
e^{i{\bf k}\cdot{\bf x}}\!- e^{i{\bf k}_B\cdot{\bf x}}) +
h.c.\big]\right]}\nonumber\\
 &&\geq -\frac{4\pi\alpha}{\delta
V}\left(\frac{3\pi c_3}{2}(\frac{3}{c_2})^{1/2}\alpha^{-3/10}\right)^2
\sum_{|{\bf
k}|<K}|{\bf k}|^{-2}\nonumber\\
&&=-\frac{108c_3^2}{c_1c_2\delta}\alpha^{8/5}.\label{explowerbound}
\end{eqnarray}
We make the choice $\delta= c_4\alpha^{-1/5}$.

The above inequality (\ref{explowerbound}) permits us to bound $H_{K}$
(still restricted to states with ${\bf x}$-support in a cube of sides with
size $L$) from below in the following way:
\begin{eqnarray}
H_{K}&\geq&(1-c_1\alpha^{-1/5}){\bf p}^2
+\sum_{B}\sum_{{\bf k}\in B}\!\Big[(1\!-c_4\alpha^{-1/5}){a}_{\bf
k}^*{a}_{\bf k}^{\phantom *}   \nonumber\\
 &+&
(\frac{4\pi\alpha}{V})^{1/2}(\frac{{a}
 _{\bf k}}{|{\bf k}|}
      e^{i{\bf k}_B\cdot{\bf x}}\! + h.c.)\Big]\!
      -\frac{108c_3^2}{c_1c_2c_4}\alpha^{9/5}.\label{block1}
\end{eqnarray}

Note that, within a single block, 
the exponential factors multiplying the $a_{\bf k}$- and $a_{\bf
k}^*$-terms are now {\it independent} of ${\bf k}$.   
With this in mind, we define,
for each block $B$, a
block annihilation operator  
\begin{equation}  
        A_B\equiv (\sum_{{\bf k}\in B}|{\bf k}|^{-2})^{-1/2} \sum_{ {\bf
        k}\in B}\frac{a_{\bf k}}{|{\bf k}|}
\end{equation}
and its corresponding adjoint $A_B^*$.  
These operators are properly normalized boson modes, i.e., 
\begin{equation}
\left[A_B, A_{B'}^* \right] = \delta_{B,B'}.
\end{equation}
Then
\begin{equation}
  \sum_{{\bf k}\in B}a_{\bf k}^*a_{\bf k}\geq A_B^*A_B^{\phantom*},
\end{equation} 
and so the operator 
 terms on the right hand side of (\ref{block1}) exceed
 the operator $H_{K}^{block}\equiv H_{K}^{block}(\{{\bf k}_B\})$
 defined by
\begin{eqnarray}
H_{K}^{block}&\equiv& (1-c_1\alpha^{-1/5}){\bf p}^2
 \!+\sum_{B}\Biggl[(1\!-\!c_4\alpha^{-1/5})A_B^*A_B^{\phantom *}\nonumber\\
&&+(\frac{4\pi\alpha}{V}\!\!\sum_{{\bf k}\in B}|{\bf k}|^{-2})^{1/2}
({A_{B}}e^{i{\bf k}_B\cdot{\bf x}}\! + h.c.)\Biggr].
\end{eqnarray}
  Referring to (\ref{Eo}) and
(\ref{block1}), we can summarize the situation: The ground state
energy $E_0$ of $H$ satisfies
\begin{equation}
E_{0}\geq
\inf_{ \Psi }\sup_{\{{\bf k}_B\}}\langle\Psi|H_{K}^{block}|\Psi
\rangle-\big(c_2+\frac{108c_3^2}{c_1c_2c_4}\big)\alpha^{9/5}\label{E1}
\end{equation}
where  the infimum  can now even be taken over $\Psi$'s with  no
restriction on their ${\bf x}$-support, and with the ${\bf k}_B$'s
chosen optimally, depending on $\Psi$.\\

{\bf IV.} In order to get a lower bound on $H_{K}^{block}$, we use the
technology of coherent states.  Let $|\xi\rangle=
\pi^{-1/2}\exp{(-\frac{1}{2}\vert\xi\vert^2+\xi A^*)}|0\rangle$ denote a
normalized coherent state for a single harmonic oscillator.  In
terms of these states one has the identity
\begin{equation}   
\exp(\mu A)\exp(\nu A^*)=\int
\exp{(\mu\xi+\nu\xi^*)}|\xi\rangle\langle\xi| d\,\xi d\,\xi^*,
\end{equation}
from which the identities for the operators
$A$ and $A^*A$ mentioned in
the introduction are readily obtained.

Let $\Psi$ be any normalized state and let $|{\bf
\xi}\rangle=\prod_B|\xi_B\rangle$ denote a tensor product of coherent
states of the block modes corresponding to the operators $\{A_B\}$.
Set $\Psi_{\bf \xi}({\bf x})= \langle{\bf \xi}|\Psi\rangle_{phonon}$,
where the inner product, $\langle \ |\  \rangle_{phonon}$, is just 
over the  phonon variables, not ${\bf x}$.  Then
\begin{equation}
\langle\Psi|H_{K}^{block}|\Psi\rangle= \int\langle\Psi_{\bf \xi}|
h_{\bf \xi}|\Psi_{\bf \xi}\rangle_{electron}\prod\nolimits_B
d \,\xi_B d\,\xi_B^*,
\label{directintegral}
\end{equation}
where the inner product in the integrand, 
$\langle \ |\  \rangle_{electron}$, is
over the electronic coordinates ${\bf x}$, and $h_{\bf \xi}$ is the
Schr\"{o}dinger operator
\begin{eqnarray}
h_{\bf \xi}&\equiv& (1\!-c_1\alpha^{-1/5}){\bf
p}^2+\sum_B\Biggl[(1\!-c_4\alpha^{-1/5})(\vert\xi_B\vert^2-1)\nonumber\\
&&+(\frac{4\pi\alpha}{V}\sum_{{\bf k}\in B}|{\bf
k}|^{-2})^{1/2}(\xi_Be^{i{\bf k}_B\cdot{\bf x}}+c.c.)\Biggr].
\end{eqnarray}

Now the supremum over ${\bf k}_B$ of (\ref{directintegral}) exceeds
what is obtained by completing the square, namely, 
\begin{equation}
\sup_{\{{\bf k}_B\}}\int \Big[(1\!-c_1\alpha^{-1/5})\langle
\Psi_{\bf \xi}|{\bf p}^2|\Psi_{\bf \xi}\rangle_{electron}
-\frac{4\pi\alpha}{(\!1\!\!-\!\!c_4\alpha^{-1/5})V}\!\sum_{B}
\sum_{{\bf k}\in B}\frac{\vert\hat{\rho}_{\bf \xi}({\bf k}_B)\vert^2}
{\vert{\bf k}\vert^{2}\hat{\rho}_{\bf \xi}({\bf 0})} \Big]
\prod\nolimits_B d\xi_B   
d\xi_B^* \  -\!N, \label{xxx}
\end{equation}
where $\hat{\rho}_{\bf \xi}({\bf
k})\!\! =\langle\Psi_{\bf \xi}|
\exp{(i{\bf k}\cdot {\bf x})}|\Psi_{\bf \xi}\rangle_{electron}$ is 
the Fourier 
transform of $\vert\Psi_{\bf \xi}\vert^2({\bf x})$. (Here, we can remove the
$ \vert {\bf k} \vert < K$ restriction on the ${\bf k}$-sum.)  We emphasize 
again that were it not for the grouping of
the phonons into blocks, then this coherent state estimate would
contain a negative constant equal to the number of phonon
modes with $|{\bf k}|<K$, 
which is  infinite in the $V\rightarrow \infty$ limit, 
instead of $N= O(\alpha^{9/5})$.

    At this point we choose
${\bf k}_B$ to be a point in $B$ where the function of ${\bf k}$
given by   
\begin{equation}
\int\frac{\vert\hat{\rho}_{\bf \xi}({\bf k})\vert^2}
{\hat{\rho}_{\bf \xi}({\bf 0})}\prod\nolimits_B d\xi_B d\xi_B^*,
\end{equation}
is minimal. 
This point depends on $\Psi$.  
With this  choice, 
(\ref{xxx}) is, in turn, larger than  
\begin{equation}
\int\Big[(1\!-c_1\alpha^{-1/5})\langle\Psi_{\bf \xi}|{\bf p}^2|
\Psi_{\bf \xi}\rangle_{electron}
-\frac{4\pi\alpha}{(\!1\!\!-\!\!c_4\alpha^{-1/5})V}\sum_{{\bf k}}
\frac{\vert\hat{\rho}_{\bf \xi}({\bf k})\vert^2}
{\vert{\bf k}\vert^{2}\hat{\rho}_{\bf \xi}({\bf 0})}\Big]
\prod\nolimits_B d\xi_B d\xi_B^*\!-\!N. \label{yyy}
\end{equation}
But the integrand of this integral is $\langle\Psi_{\bf \xi}|
\Psi_{\bf \xi}\rangle_{electron}\times{\cal E}(\Psi_{\bf \xi}/(\langle
\Psi_{\bf \xi}|\Psi_{\bf \xi}\rangle_{electron}^{1/2}))$ 
, where ${\cal E}(\cdot)$ is just the Pekar functional (\ref{Pekar}),
 (but with coefficients altered by $O(\alpha^{-1/5})$); the ${\bf
 k}$-sum is the Fourier series for the Coulomb self-energy term of
 ${\cal E}$.  Thus, the integral (\ref{yyy}) exceeds
 $(1\!-\!c_1\alpha^{-1/5})^{-1}(1\!-\!c_4\alpha^{-1/5})^{-2}E_P(\alpha)$,
 with $E_P(\alpha)$ being the Pekar minimum energy of (\ref{Pekar}).
 
 Taking into account the altered coefficients in the Pekar functional, the
 lower bound (\ref{E1}), and the definition of $N$, we obtain
\begin{equation}
E_{0}(\alpha)\geq E_{P}(\alpha)
-\big(c_2\!+\!\frac{108c_3^2}{c_1c_2c_4}\!+\!\frac{2^{11}}
{3\pi^{2}c_1^3c_3^3}+c_5\big)\alpha^{9/5}\!-\!o(\alpha^{9/5})\label{E2}
\end{equation}
where $E_P(\alpha)= -c_P\alpha^2$, $c_P= 0.109$,\cite{GL},
 and where
$c_5=(c_1+2c_4)c_P$ accounts for the altered coefficients in the
variational principle.  This completes the proof of the lower bound.
Optimizing the coefficients, we find that the $\alpha^{9/5}$-error term
 in (\ref{E2})
is no greater than $3.822\alpha^{9/5}$.

This work was partially supported by NSF grant 
PHY95--13072(E.H.L.).

\noindent{\bf Erratum}\\

We are grateful to Professor Andrey V. Soldatov of the Moscow Steklov
Mathematical Institute for calling our attention to an error in our
paper.  The commutator inequality (8) in our step {\bf I},
namely $|k_j||\langle a_{\bf
  k}e^{i{\bf k}\cdot{\bf x}}\rangle|\leq 2\langle
p_j^2\rangle^{1/2}\langle a_{\bf k}^*a_{\bf k}^{\phantom
  *}\rangle^{1/2}$, is not correct.  Rather, the right side
of this inequality should be $\langle p_j^2\rangle^{1/2}(\langle
a_{\bf k}^*a_{\bf k}^{\phantom *}\rangle^{1/2}+\langle a_{\bf
  k}^{\phantom*}a_{\bf k}^{*}\rangle^{1/2})$ or a related expression.  
The extra factor
$\langle a_{\bf k}^{\phantom*}a_{\bf k}^{*}\rangle^{1/2}$ with the
$a_{\bf k}^{\phantom*}$ and $a_{\bf k}^{*}$ not in normal order
generates uncontrolled mischief with, for example,  the right side of
the ultraviolet bound (10) containing an additional term $\sum_{|{\bf
    k}| \geq K}1/2 = \infty$.

The situation is remedied with the help of the method introduced
by Lieb and Yamazaki [14] to obtain the previous rigorous
lower bound on the polaron energy. Our main result, (31), is still valid. Indeed, it is improved slightly. 

Define the (vector) operator ${\bf Z}=
(Z_1, Z_2, Z_3)$ with components
\begin{equation}
  Z_j= (\frac{4\pi\alpha}{V})^{1/2}\sum_{|{\bf k}|\geq
K} k_j \frac{{a}_{\bf k}}{|{\bf k}|^3}e^{i{\bf k}\cdot{\bf
x}},\hspace{.2 in} j= 1,2,3.
\end{equation}
Then the commutator estimate (8) is replaced by 
\begin{eqnarray}
  -(\frac{4\pi\alpha}{V})^{1/2}\sum_{|{\bf k}|\geq
    K}\left[\langle\frac{{a}_{\bf k}}{|{\bf k}|}e^{i{\bf k}\cdot{\bf
      x}}\rangle +c.c.\right] &\equiv& -\sum_{j}\langle [p_j,
  Z_j^{\phantom *}-Z_j^{*}] \rangle \nonumber \\ \leq 2\langle {\bf
    p}^2 \rangle ^{1/2} \langle -({\bf Z}-{\bf Z^*})^2 \rangle ^{1/2}
  &\leq& 2\langle {\bf p}^2 \rangle ^{1/2}\langle2({\bf Z}^*{\bf
    Z}+{\bf Z}{\bf Z}^*)\rangle^{1/2}\nonumber\\ &\leq&
  \varepsilon\langle {\bf p}^2\rangle
  +\frac{2}{\varepsilon}\langle{\bf Z}^*{\bf Z}+{\bf Z}{\bf
    Z}^*\rangle.
\end{eqnarray}
Now, each component $Z_j$ can be thought of as a single (unnormalized)
oscillator mode having commutator with its adjoint, $[Z_j^{\phantom
  *}, Z_j^*]=(4\pi\alpha /V)\sum_{|{\bf k}|\geq K}k_j^2|{\bf
    k}|^{-6}\rightarrow 2\alpha /3\pi K$; moreover, $Z_i^{\phantom *}$ and
$Z_j^*$ commute for $i\neq j$ (i.e., these modes are orthogonal).
Using these facts, we have that 
\begin{eqnarray}
  \frac{2}{\varepsilon}\langle{\bf Z}^*{\bf Z}+{\bf Z}{\bf Z}^*\rangle
  &=& \frac{4}{\varepsilon}\langle {\bf Z}^*{\bf Z}\rangle +
  \frac{2}{\varepsilon}\left(\frac{2\alpha}{\pi K}\right)\nonumber \\ 
  &\leq& \sum_{|{\bf k}|\geq K}\langle a_{ \bf k}^*a_{\bf k}^{\phantom
    *}\rangle+ 3/2
\end{eqnarray}
if we choose $\varepsilon = 8\alpha /3\pi K $, which is smaller and
better by a factor $1/3$ from the $\varepsilon$ in the article.
 Here we have employed an
orthogonal rotation of coordinates bringing $\sum_{|{\bf
    k}|\geq K}a_{\bf k}^*a_{\bf k}^{\phantom *} $ into a form
$(4/\varepsilon){\bf Z}^*{\bf Z}+ $non-negative operators. (Compare
Eqs.(21,22) of the article.) Combining these inequalities, we obtain
\begin{equation}
  -(\frac{4\pi\alpha}{V})^{1/2}\sum_{|{\bf k}|\geq
    K}\left[\langle\frac{{a}_{\bf k}}{|{\bf k}|}e^{i{\bf k}\cdot{\bf
      x}}\rangle +c.c.\right] \leq \varepsilon\langle{\bf
    p}^2\rangle+\sum_{|{\bf k}|\geq K}\langle a_{\bf k}^*a_{\bf
    k}^{\phantom *}\rangle+ 3/2.
\end{equation}

This last inequality is our replacement for the ultraviolet bound
(10).  It follows that $H\geq H_K-3/2$ where $H_K$ is as in
Eq.(11), but with the coefficient of ${\bf p}^2$ given by
$(1-8\alpha/3\pi K)$ rather than $(1-8\alpha/\pi K)$.  With  the
choice $K=
8\alpha/3\pi$, inequality (13) becomes $H\geq
-(16\alpha^2/3\pi^2)-3/2$, a bound at least
consistent with a known {\it upper} bound for the ground state energy
linear in $\alpha$.

The remainder of the article is an analysis of $H_K$ and needs only
minor modification.  The coefficient of ${\bf
  p}^2$ in Eqs.(19,23,27,28,30) should be $(1-c_1\alpha^{-1/5}/3)$
and, at the end of the article, $c_5= (c_1/3+2c_4)c_P$.  Due to the
smaller value of $\varepsilon$ defined above, our estimate
on the
coefficient of $\alpha^{9/5}$ in (31) is slightly improved to 2.337,
rather than 3.822 as reported.  Of course, our lower bound for the
ground state energy is decreased merely by the constant $-3/2$, which
is unimportant on a scale of $\alpha^{9/5}$.   
\end{document}